\documentclass[11pt]{article}
\usepackage{moriond}

\usepackage{amsmath,amssymb}

\def\Journal#1#2#3#4{{#1} {\bf #2}, #3 (#4)}

\def\NPB{{\em Nucl. Phys.} B}

\def\PRL{\em Phys. Rev. Lett.}
\def\PRD{{\em Phys. Rev.} D}

\def\be{\begin{equation}}
\def\ee{\end{equation}}
\def\bea{\begin{eqnarray}}
\def\eea{\end{eqnarray}}

\renewcommand{\eqref}[1]{Eq.~\ref{#1}}
\newcommand{\figref}[1]{Figure~\ref{#1}}

\def\JHEP{\em JHEP}

\def\SMSt{\Delta m_{\tilde\ell}(\tilde e_L, \tilde\mu_L) / {m_{\tilde\ell}}}

\newcommand{\Photo}{}

\begin{document}
\vspace*{-2cm}
\begin{flushright}
 CFTP/13-014 \\
 PCCF-RI-13-04
\end{flushright}
\vspace*{4cm}
\title{The high-scale SUSY seesaw: LHC vs low energy}

\author{A. J. R. Figueiredo}

\address{CFTP, Dep.\ de F\'isica -- IST, Av.\ Rovisco Pais, 1, 1049-001 Lisboa, Portugal\\ \& LPC, Laboratoire de Physique Corpusculaire, CNRS/IN2P3 -- UMR 6533,\\ Campus des C\'ezeaux, 24 Av. des Landais, F-63177 Aubi\`ere Cedex, France}

\maketitle\abstracts{
	In this contribution we outline the correlation between intergenerational slepton mass 
	splittings and low energy lepton flavour violation in supersymmetric type-I and type-III 
	seesaws, and illustrate how the combination of these two sets of observables could 
	strengthen or disfavour a high-scale seesaw as the explanation of neutrino masses and 
	mixings. This contribution summarises part of the analysis presented 
	in\cite{Abada:2010kj,Abada:2011mg}.}
	
\section{Introduction}

	Under the assumption that the seesaw is the unique source of lepton flavour violation (LFV) 
	and that slepton masses are flavour universal at an energy scale higher than that of the seesaw, 
	intergenerational slepton mass splittings at low energy (i.e.\ $m_{\tilde \ell_i}$ -- $m_{\tilde \ell_j}$) 
	are tightly correlated with low energy charged LFV (cLFV). 
	This is a direct consequence of their common origin in radiatively 
	generated slepton flavour mixing\cite{Borzumati:1986qx}. 
	For the illustrative examples of the supersymmetric (SUSY) fermionic seesaws, we will show that 
	the aforementioned correlations have the potential of probing these high-scale SUSY seesaws.

\section{Supersymmetric fermionic seesaws} \label{sec.models}

	We will focus on seesaw realisations that rely on the exchange of heavy fermions $F$ 
	to generate neutrino masses and mixings. At the fundamental level, the terms in the 
	Lagrangian density responsible for neutrino masses read 
	\be
		Y^\nu_{ij} F_i L_j H_u - \frac{1}{2} M_{ij} F_i F_j \,,
	\ee
	where $i$,$j$ are generation indices. If $L_i H_u$ is a weak singlet (triplet), $F_i$ must be a weak 
	singlet (triplet) and we have the so-called type-I (type-III) seesaw mechanism. At low energies 
	the fundamental Lagrangian is matched by an effective Lagrangian that has 
	a dimension-5 operator resulting from integrating out the heavy fermions $F_i$ and which, 
	after electroweak symmetry breaking (EWSB), generates neutrino masses 
	and mixings. Approximately, we have 
	\be
		m_\nu \simeq -v^2_u \,\, {Y^\nu}^T \, M^{-1} \, Y^\nu \,,
	\ee
	where $v_u$ is the vacuum expectation value of $H_u$. This relation suggests that a convenient way 
	of parametrizing the neutrino 
	Yukawa couplings $Y^\nu$ at the seesaw scale, while at the same time allowing 
	to accommodate neutrino data, is given by\cite{Casas:2001sr} 
	\be
		Y^\nu = \frac{i}{v_u} \sqrt{M^\text{diag}} \, R \, \sqrt{m^\text{diag}_\nu} \, {U^\text{MNS}}^\dagger \,, 
		\label{eq:CasasIbarra}
	\ee
	where $U^\text{MNS}$ is the neutrino mixing matrix and $R$ is a complex orthogonal matrix that 
	encodes mixings among heavy fermions; \eqref{eq:CasasIbarra} is cast in a basis in which 
	the charged lepton Yukawa couplings ($Y^l$) and $M$ are diagonal in generation space. 
	Hereafter, we will work in this basis. 

	In order to preserve the attractive features of SUSY gauge coupling unification, we will embed the heavy 
	fermion triplets in a SU(5) grand unified theory (GUT). The lowest dimensional 
	representation that contains a weak triplet that is a singlet under SU(3)$\otimes$U(1) is a 
	$\mathbf{24}$-plet whose decomposition under SU(3)$\otimes$SU(2)$\otimes$U(1) 
	reads 
	\be
		\mathbf{24} = \left(1,3,0\right) \oplus \left(1,1,0\right) \oplus \text{coloured fields} \,. 
	\ee
	Thus, the type-III implementation by means of $\mathbf{24}$-plets is accompanied by a 
	type-I seesaw. 
	
	Our analysis for the type-I seesaw will be conducted in the minimal supersymmetric SM (MSSM) 
	extended by three 
	heavy right-handed (RH) neutrino superfields while in the $\mathbf{24}$-plet seesaw we will 
	consider the supersymmetrized version of the Georgi-Glashow SU(5) model supplemented 
	by three generations of $\mathbf{24}$-plet superfields. For details on the extended superpotential and 
	SUSY soft-breaking sector we refer to the works\cite{Abada:2010kj,Abada:2011mg} 
	upon which this contribution is based. 
	 
	In addition, we will impose at the GUT scale ($M_\text{GUT} \sim 10^{16}$ GeV) minimal 
	supergravity (mSUGRA) 
	inspired universality conditions for the SUSY soft-breaking sector of each model: a common mass 
	$m_0$ for the scalars (including the scalar partners of the heavy fermions); a common 
	mass for the gauginos, $M_{1/2}$; trilinear couplings given by 
	$A^{u,d,l,\nu} = A_0 Y^{u,d,l,\nu}$ (where $Y^\nu \equiv Y^{24}$ in the case of the 
	$\mathbf{24}$-plet seesaw\cite{Abada:2011mg}). The values of $Y^{u,d}$ at $M_\text{GUT}$ 
	are fitted to give the low energy quark masses and mixings, while $Y^l$ and 
	$m^\text{diag}_\nu$ in \eqref{eq:CasasIbarra} (with an additional overall factor of $\sqrt{5/4}$ for 
	the $\mathbf{24}$-plet seesaw case) are fitted to reproduce 
	the low energy lepton masses and mixings. In doing this we are taking as input for $U^\text{MNS}$ 
	in \eqref{eq:CasasIbarra} the low energy neutrino mixing matrix, exploiting that the running of 
	the mixing angles is negligible. 
	
	The running from $M_\text{GUT}$ down to the seesaw scale will induce flavour mixing 
	\cite{Borzumati:1986qx} in the approximately flavour conserving slepton soft breaking terms 
	due to the non-trivial flavour structure of $Y^\nu$ (implied by neutrino mixing and by possible 
	heavy fermion mixing). This effect is more pronounced in the soft breaking terms involving slepton 
	doublets since these have local interactions with the heavy fermions. 
        At leading order, the flavour mixing induced by these radiative corrections is given by 
        \bea
		&& \left( \Delta m^2_{\tilde L} \right)_{ij} = -a_F \frac{1}{8\pi^2} 
			\left( 3 \, m^2_0 + A^2_0 \right) 
			\left( {Y^\nu}^\dagger L Y^\nu \right)_{ij} \,,\\
		&& \left( \Delta A^l \right)_{ij} = -a_F \frac{3}{16\pi^2} A_0 Y^l_{ij} 
			\left( {Y^\nu}^\dagger L Y^\nu \right)_{ij} \,;\, 
		L_{kl} \equiv \log\left(\frac{M_{\text{GUT}}}{M_k}\right) \delta_{kl} \,,
	\eea
	where $a_F = 1$ for the singlet seesaw and $a_F = 9/5$ for the $\mathbf{24}$-plet seesaw. 
	The amount of flavour violation in the slepton sector is encoded in 
	$\left( {Y^\nu}^\dagger L Y^\nu \right)_{ij}$ which, as made explicit in 
	\eqref{eq:CasasIbarra}, is related to heavy fermion masses and mixings, as well 
	as to neutrino masses and mixings. If the seesaw scale is $\sim 10^{15}$ GeV, 
	the neutrino Yukawa couplings are of $\mathcal{O}(1)$ and the radiative 
	corrections will yield sizeable slepton flavour mixing, with potentially observable effects.

\subsection{Low energy cLFV observables}

	An effect of sizeable slepton flavour mixing are potentially large new SUSY contributions 
	to cLFV observables (through sneutrino-chargino and slepton-neutralino loops). For the 
	case of cLFV radiative decays $\ell_i \to \ell_j \gamma$, a simple illustrative expression 
	can be obtained using the leading-logarithm approximation 
	\be
		\frac{\text{BR($\ell_i\to\ell_j\gamma$)}}{\text{BR($\ell_i\to\ell_j\nu_i\bar\nu_j$)}} = 
			\frac{\alpha^3 \tan^2\beta}{G^2_F m^8_\text{SUSY}} 
			\left| a_F \frac{1}{8 \pi^2} \left( 3 \, m^2_0 + A^2_0 \right) \left( {Y^\nu}^\dagger L Y^\nu \right)_{ij} \right|^2 \,.
	\ee

\subsection{Intergenerational slepton mass splittings}

	Under the assumption that SUSY breaking generates flavour universal slepton soft-breaking masses, 
	intergenerational slepton mass splittings may arise from left-right mixing 
	after EWSB and/or renormalisation group (RG) running. In the MSSM both of them are 
	proportional to the only source of flavour non-universality, $Y^l$. Neglecting 
	RG corrections as source of intergenerational slepton mass splittings, one finds that the 
	$\tilde e_L$,$\tilde \mu_L$ mass difference normalised to their average mass, $m_{\tilde\ell}$, 
	is approximately given by 
	\be
		\frac{\Delta m_{\tilde\ell}}{m_{\tilde\ell}}(\tilde e_L,\tilde\mu_L) \approx \frac{m^2_\mu}{2 m^2_{\tilde\ell}} 
			\left| \frac{\left( A_0 - \mu\tan\beta \right)^2}{0.35 M^2_{1/2} + M^2_Z \cos 2\beta \left( -1/2 + 2 \sin^2\theta_W \right)} \right| \,.
	\ee
	Due to the smallness of $Y^\mu$, the mass splitting between the first two slepton generations 
	in the MSSM with mSUGRA boundary conditions was found\cite{Abada:2010kj} to be 
	always below the $0.01$\% level. 
	In the presence of radiative corrections induced by the seesaw (which we 
	take to be sufficiently heavy so that $Y^\nu \sim 1$), and assuming $(\Delta m^2_{\tilde L})_{ij}$ as 
	the dominant off-diagonal entry, the mass splitting between the mostly left-handed (LH) 
	$i$ and $j$ sleptons is approximately 
	\be 
		\frac{\Delta m_{\tilde\ell}}{m_{\tilde\ell}}(\tilde\ell_i,\tilde\ell_j) \approx 
			\left| \frac{(\Delta m^2_{\tilde L})_{ij}}{m^2_{\tilde L}} \right| \propto 
			\sqrt{\text{BR($\ell_i\to\ell_j\gamma$)}} \,. \label{eq:SMS_BR}
	\ee
	\eqref{eq:SMS_BR} illustrates in a simple way the correlation of these high- and low-energy 
	observables in the case of a single source of flavour violation in the lepton sector, $Y^\nu$. 
	
\section{Illustrative results and discussion} \label{sec.results}

\begin{figure}
		\begin{minipage}{0.5\linewidth}
		\centerline{\includegraphics[width=1\linewidth,draft=false]{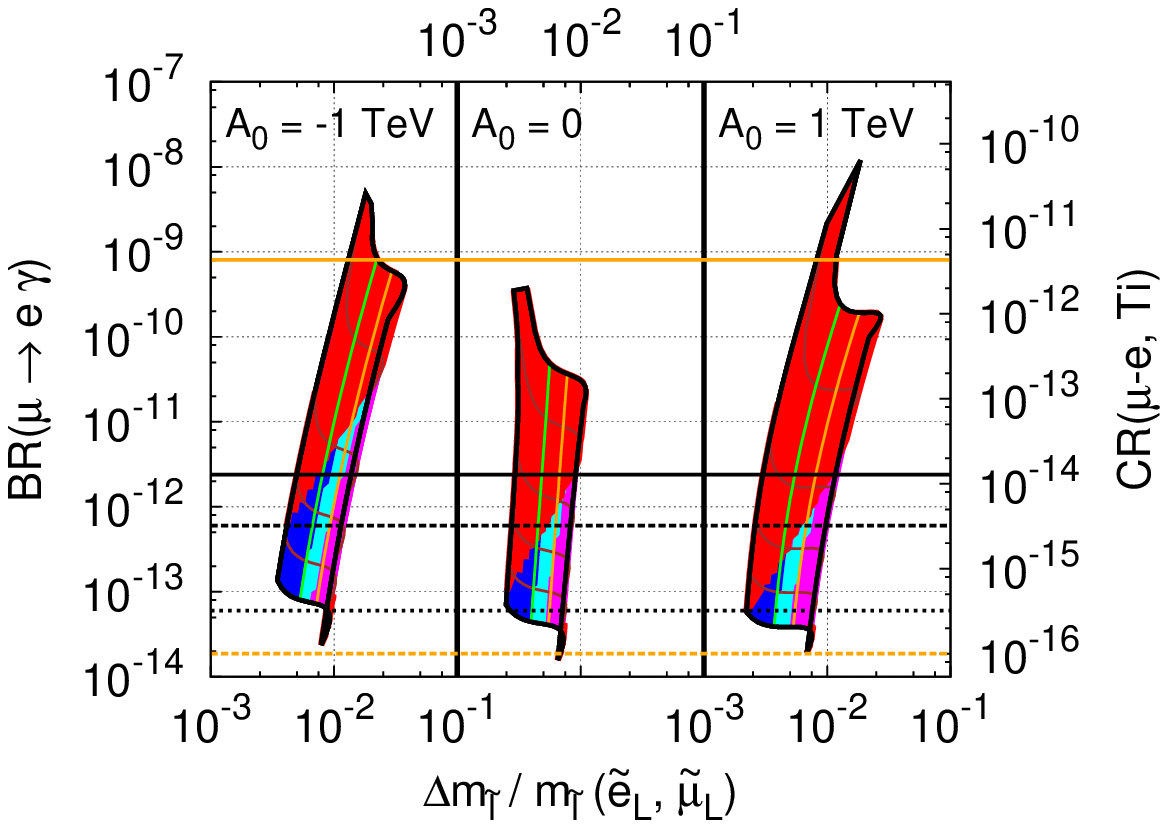}}
		\end{minipage}
		~~~~~~~
		\begin{minipage}{0.5\linewidth}
		\centerline{\includegraphics[width=1.25\linewidth,draft=false]{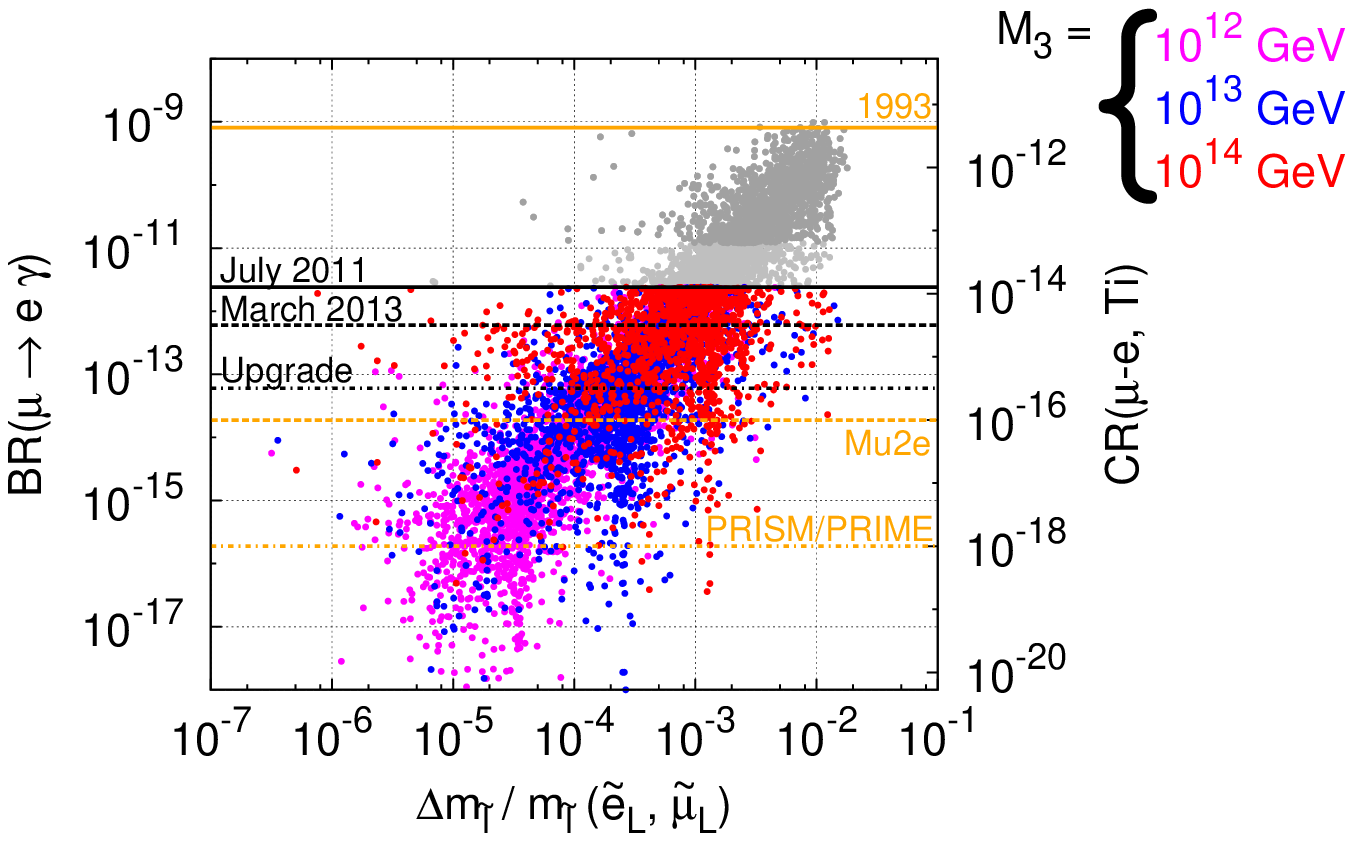}}
		\end{minipage}\vspace{-10pt}
		\caption{Correlation between BR($\mu\to e\gamma$) and the mass difference 
			$\tilde e_L$ -- $\tilde \mu_L$, normalised to $\langle \tilde e_L, \tilde \mu_L \rangle$. 
			On the secondary right y-axis we show the corresponding predictions for 
			CR($\mu$-$e$, Ti). 
			Black (orange) horizontal lines correspond to current bounds and future sensitivities 
			on $\mu\to e\gamma$ ($\mu$-$e$ conversion in Ti). 
			On the left: $\mathbf{24}$-plet seesaw with a degenerate $\mathbf{24}$-plet spectrum. 
			We have set $m_0 = 100$ GeV, $\tan\beta = 10$ and $\theta_{13} = 0.1^\circ$, 
			and varied the $\mathbf{24}$-plet scale and $M_{1/2}$ in the ranges 
			$\left[10^{13},9\times 10^{14}\right]$ GeV and $\left[1.5,6\right]$ TeV, respectively. 
			Red regions are excluded due to having a too light SM-like scalar boson. 
			Magenta regions have a charged lightest supersymmetric particle, while blue (cyan) regions 
			have a wino-like neutralino lighter (heavier) than the LH sleptons. 
			On the right: type-I SUSY seesaw with a randomly varied RH neutrino mixing 
			matrix: $|\theta_i| \leq \pi$ and $\arg\theta_i \in \left[-\pi,\pi\right]$. 
			We have taken benchmark point P5-HM1{\protect\footnotemark}, $\theta_{13} = 0.1^\circ$ and 
			hierarchical RH neutrinos: $M_{1(2)} = 10^{10(11)}$ GeV and three values of $M_3$ 
			(coloured regions). Dark- and light-grey points are excluded by upper-limits on 
			BR($\mu\to e\gamma$) prior to March 2013. Light-grey points were allowed when this 
			analysis was first done. Points above the dashed black line are excluded by the latest 
			results reported by the MEG collaboration.}
		\label{fig}
	\end{figure}

	In the left panel of \figref{fig} we display the correlation between 
	slepton mass splittings and two low energy cLFV observables, namely 
	BR($\mu\to e\gamma$) and CR($\mu$-$e$, Ti), for the $\mathbf{24}$-plet seesaw. 
	 ``Horizontal'' isolines correspond to constant values of $M_{1/2}$ with the $\mathbf{24}$-plet mass 
	 scale ($M_{24}$) increasing from left to right. For further details see\cite{Abada:2011mg}. 		
	The decrease 
	in BR($\mu\to e\gamma$) with increasing $M_{24}$ along ``horizontal'' isolines 
	is due to the way the $\mathbf{24}$-plets affect the SUSY spectrum: sleptons become lighter 
	as $M_{24}$ gets farther from $M_\text{GUT}$. The balance between reducing slepton flavour 
	mixing and alleviating the loop mass suppression is such that the latter dominates slepton 
	mediated $\mu$-$e$ transitions. In contrast, $\SMSt$ increases with 
	increasing $M_{24}$. 
 
	In the right panel of \figref{fig}, with a different underlying scan, 
	we illustrate similar correlations for the type-I seesaw case. 
	Leading to the figure we have performed a random scan over 
	the RH neutrino mixing matrix. These results confirm the correlation 
	between the two sets of observables and the dependence displayed in 
	\eqref{eq:SMS_BR}. 
	
	In the $\mathbf{24}$-plet seesaw, slepton mass splittings between the first two generations 
	are expected\cite{Abada:2011mg} to lie in the $0.1$-$1$\% level for 
	a SUSY spectrum within LHC range and for the entire viable 
	range of the $\mathbf{24}$-plet mass scale. In contrast, for the type-I seesaw 
	we have found\cite{Abada:2010kj} that mass splittings above the $0.1$\%  
	level restrict the RH neutrino scale to lie above $10^{13}$ GeV. 
	
	Assuming that sleptons are discovered at the LHC, their mass splittings can be compared to the 
	results available on cLFV observables and interpreted under the light of both a high-scale 
	SUSY model and its seesaw extension. If the high-scale SUSY model can be inferred 
	from a subset of the data (independently from a possible seesaw extension), then by 
	combining the information on low energy cLFV observables with intergenerational slepton 
	mass splittings the hypothesis of a particular seesaw extension of a high-scale SUSY model can be 
	strengthened, if it accommodates both sets of observables without ad hoc sources of LFV or of 
	intergenerational slepton mass differences; likewise it can be disfavoured, if accommodating both 
	sets of observables is difficult. 

	Although in our analysis we have set $\theta_{13} = 0.1^\circ$, we stress, however, that the results 
	shown in \figref{fig} are only mildly sensitive\cite{Abada:2010kj} to the value of 
	$\theta_{13}$ and little changes would appear from considering 
	$\theta_{13} \sim 9^\circ$ \cite{GonzalezGarcia:2012sz} (as now experimentally measured). 
	Indeed, slepton flavour mixing generated by a degenerate 
	heavy fermion spectrum with $R = 1$, as taken in the scan leading to the left panel results, 
	has a small dependence on the size of $\theta_{13}$. Moreover, a scan over the parameter space of 
	the heavy fermion mixing matrix, as 
	was done for the right panel, hides the dependence of slepton flavour mixing on the 
	precise value of $\theta_{13}$. 
	
	\footnotetext{This benchmark point is now excluded by the LHC results on the SM-like scalar 
			boson mass; nevertheless, a P5-HM1 variant with large $-A_0$ is able to lift the mass so as 
			to be compatible with the LHC without significantly affecting the conclusions, but at the expense 
			of shifting the points upwards along the correlation slope.}

\section{Conclusion} \label{sec.conclusions}

	We have shown that if sleptons are discovered and their masses reconstructed to a tentative  
	accuracy of $0.1\%$\cite{Bachacou:1999zb} for first- and second-generation 
	sleptons, the current upper-bounds and future results of low energy 
	cLFV experiments have the potential to strengthen or disfavour the high-scale SUSY 
	seesaw explanation of neutrino masses and mixings. 

\section*{Acknowledgements}

	This work has been supported by {\it Funda\c c\~ao para a Ci\^encia e a 
	Tecnologia} (FCT) through the fellowship SFRH/BD/64666/2009 and grants 
	CFTP-FCT UNIT 777 and CERN/FP/123580/2011.

\end{document}